\documentclass[aps,prc,reprint,superscriptaddress,showpacs,amsmath,amssymb]{revtex4-1}
\usepackage{graphicx}
\usepackage{hyperref}
\usepackage{float}
\usepackage{dcolumn}
\usepackage{bm}
\usepackage{color}
\begin{document}

\title{Particle number conserving BCS approach in the relativistic mean field model and its application to $^{32-74}$Ca}

\author{Rong An}
\affiliation{School of Physics and Nuclear Energy Engineering,
Beihang University, Beijing 100191, China}
%\thanks{E-mail address: freeanrong@163.com/rongan@buaa.edu.cn}

\author{Lisheng Geng}
\email[E-mail: ]{lisheng.geng@buaa.edu.cn}
\affiliation{School of Physics and
Nuclear Energy Engineering \&
Beijing Key Laboratory of Advanced Nuclear Materials and Physics,  Beihang University, Beijing 100191, China\&
Beijing Advanced Innovation Center for Big Date-based Precision Medicine, Beihang University, Beijing100191, China
}

\author{Shisheng Zhang}
\email[E-mail: ]{zss76@buaa.edu.cn}
\affiliation{School of Physics and Nuclear Energy Engineering,
Beihang University, Beijing 100191, China}

\author{Lang Liu}
\email[E-mail: ]{liulang@jiangnan.edu.cn}
\affiliation{School of Science, Jiangnan University, Wuxi 214122, China}

\date{\today}

\begin{abstract}

A particle number conserving BCS approach (FBCS) is formulated in  the relativistic mean
field (RMF) model. It is shown that the so-obtained RMF+FBCS model can describe the weak
pairing limit. We calculate the ground-state properties of the calcium isotopes $^{32-74}$Ca and
compare the results with those obtained from the usual RMF+BCS model. Although the results
are quite similar to each other, we observe an interesting phenomenon, i.e.,  for $^{54}$Ca, the FBCS approach
can enhance the occupation probability of the $2p_{1/2}$ single particle level and slightly increases its radius, compared with the RMF+BCS model.
This leads to an unusual scenario that although $^{54}$Ca
is more bound with a spherical configuration but the corresponding size is not the most compact one.
We anticipate that such a phenomenon might happen for other neutron rich nuclei
and should be checked by further more systematic studies.
\end{abstract}

%\pacs{25.70.Jj,24.10.-i}

\maketitle
\section{INTRODUCTION}
In recent years, studies of exotic nuclei with large isospin ratios have become the
forefront of nuclear physics both theoretically and experimentally (see, e.g., Refs.~\cite{
Tanihata:2013jwa,Meng:2015hta} and references cited therein). This brings great challenges to existing nuclear structure
models for a reliable understanding, interpretation and prediction of new experimental phenomena. Two
of  the crucial theoretical issues (at least in mean-field models) are:
(i) a proper description of the continuum; and (ii) a reliable treatment of
the residual pairing correlation. Both subjects have been extensively studied ~\cite{Dobaczewski:1995bf, PhysRevLett.77.3963, Poschl:1997ky, Lalazissis:1997mg,Meng19983,Sandulescu:1998dm,PhysRevC.66.024311, PhysRevC.68.054323,Geng:2003pk,PhysRevC.84.024311,Zhang2013,Tian:2017fbc}. The pairing
correlation has long been known to be essential to describe many experimental observables,
such as moments of inertia, level densities, and energies of the lowest-lying excited states
~\cite{P.Ring,A.Bohr}. It plays an more important role
 for weakly bound nuclei, where it is the only attractive force responsible for their
existence in mean-field models.

Conventionally, the pairing correlation can be treated either by the Bardeen-Cooper-Schrieffer
(BCS)~\cite{Bardeen:1957kj,Bardeen:1957mv,cooper2011bcs} method or by the Bogoliubov transformation~\cite{Bogolyubov:1958se}.
In earlier days, it has been realized that these methods originally developed for macroscopic
systems result in spurious sharp phase transitions from normal states to superfluid states~\cite{cooper2011bcs},
which have never been observed in experiments.
The sharp phase transitions are due to the  breaking of  particle number conservation in finite nuclei and the fact that
only the expectation value of the particle number operator is fixed.
In a macroscopic system, it can be safely ignored since the particle number is large enough. However,
in a microscopic system, such as atomic nuclei, it can lead to spurious effects which should be carefully studied.
These early findings have led to a lot of efforts in developing alternative approaches which improve the treatment of the pairing correlation.
%such as the Lipkin-Nogami approach~\cite{Nogami:1964zz}
%and the FBCS method~\cite{Dietrich:1964zz}.
The generally accepted approach to restore the broken gauge symmetry of particle number is the projection technique,
see e.g., Refs.~\cite{Dietrich:1964zz, Dean:2002zx, Sheikh200071, PhysRevC-66-044318, Anguiano2001467, Anguiano-2002nb}.
%There arealso the so-called exact methods~\cite{Zelevinsky2004299}.
The differences among the various treatments
have been studied in much detail.
It is found that most treatments are quite similar to each other in the strong pairing limit,
while only the variation after projection methods can properly describe the weak pairing limit.

A pairing method which conserves the gauge symmetry of particle number is particularly
desirable for weakly bound nuclei because (i) the pairing correlation is the  sole force to bind
the nucleus and (ii) only a few single particle levels around the Fermi  surface are important for the pairing
correlation \cite{PhysRevC.68.054323}. Therefore, it will be very interesting to formulate such
a method within a reliable mean-field model and study its impact on relevant physical quantities.

In the present work, we formulate the FBCS method~\cite{Dietrich:1964zz} in
the relativistic mean field model, one of the two most successful mean field models ~\cite{Bender:2003jk}.
To our knowledge, so far, only the Lipkin-Nogami BCS method~\cite{PhysRevC.73.034308,PhysRevC.74.064309}, the exact approach~\cite{Chen:2013jsa},  and
the Shell-model-like approach (SLAP)~\cite{JOUR,CLZ04102} have been explored in the relativistic mean field model.

This paper is organized as follows. In Section II, we briefly review the relativistic
mean field model. In Section III, we introduce the FBCS method and its implementation in the relativistic mean field model.
In Section IV, we explain how the residuum integrals are solved numerically.
In Section V, we study the general features of the RMF+FBCS model
by comparing its results with those of the RMF+BCS model.
In Section VI, we  check how well the ground-state properties of the calcium isotopes can be described
 by these two different approaches.
 Finally, we summarize and point out  possible future  extensions in Section VII.

\section{THE RELATIVISTIC MEAN FIELD MODEL}
The basic assumptions made in the relativistic mean field model is that
the nucleons  are point-like Dirac fermions and their interactions are mediated via
meson exchanges. One can then write down the relativistic Lagrangian densities for both
nucleons and mesons as well as photons. Adopting the so-called mean-field and
no-sea approximations, one then solves the coupled equations self-consistently. For a more detailed
explanation of the RMF model and the recent devlopments, see, e.g., Refs.~\cite{Walecka:1974qa,Reinhard:1989zi,Serot:1984ey,Ring:1996qi,Meng:2005jv,Vretenar:2005zz,
jie2016relativistic}.

The Lagrangian density used in this study has the following form:
\begin{eqnarray}
\mathcal{L}&=&\bar{\psi}[i\gamma^\mu\partial_\mu-M-g_\sigma\sigma
-\gamma^\mu(g_\omega\omega_\mu+g_\rho\vec
{\tau}\cdotp\vec{\rho}_{\mu}\\\nonumber
&+&e\frac{1-\tau_3}{2}A_\mu)
-\frac{f_\pi}{m_\pi}\gamma_5\gamma^\mu\partial_\mu
%\overrightarrow{\pi}\cdotp\overrightarrow{\tau}]\psi\\\nonumber
\vec{\pi}\cdotp\vec{\tau}]\psi\\\nonumber
&+&\frac{1}{2}\partial^\mu\sigma\partial_\mu\sigma-\frac{1}{2}m_\sigma^2\sigma^2
-\frac{1}{3}g_{2}\sigma^{3}-\frac{1}{4}g_{3}\sigma^{4}\\\nonumber
&-&\frac{1}{4}\Omega^{\mu\nu}\Omega_{\mu\nu}+\frac{1}{2}m_{\omega}^2\omega_\mu\omega^\mu
+\frac{1}{4}c_{3}(\omega^{\mu}\omega_{\mu})^{2}\\\nonumber
&-&\frac{1}{4}\vec{R}_{\mu\nu}\cdotp\vec{R}^{\mu\nu}+\frac{1}{2}m_\rho^2\vec{\rho}^\mu\cdotp\vec{\rho}_\mu
+\frac{1}{4}d_{3}(\vec{\rho}^{\mu}\vec{\rho}_{\mu})^{2}\\\nonumber
&+&\frac{1}{2}\partial_\mu\vec{\pi}\cdotp\partial^\mu\vec{\pi}-\frac{1}{2}m_\pi^2\vec{\pi}\cdotp\vec{\pi}\\\nonumber
&-&\frac{1}{4}F^{\mu\nu}F_{\mu\nu},
\end{eqnarray}
where all symbols have their usual meanings. The corresponding Dirac equation
for the nucleons and Klein-Gordon equations
for the mesons and photon, obtained with the mean-field and the
no-sea approximation, are solved by the expansion method with the harmonic
oscillator basis~\cite{Gambhir:1989mp,Ring:1997tc,Geng:2003pk}. In the present
work, 12 shells are used to expand the fermi fields and 20 shells for
the meson fields. The mean-field effective force used is
NL3 \cite{PhysRevC.55.540}, and we found that using other effective forces
such as TM1~\cite{SUGAHARA1994557} and PK1~\cite{PhysRevC.69.034319} do not
change essentially any of our conclusions.

\section{THE FBCS METHOD}
The FBCS method has been known for a long time~\cite{Dietrich:1964zz}, but to our knowledge, it has not
been applied in the relativistic mean field model in a self-consistent manner. Here we briefly describe some
essential ingredients of this approach. A detailed derivation can be found in Ref.~\cite{Dietrich:1964zz}.
In order to simplify the final FBCS equations and also to simplify the derivation, we adopt the notion
of the ``residuum integrals'' introduced by Dietrich, Mang and Pradal~\cite{Dietrich:1964zz}. Introducing
a complex variable $z = e^{i\psi}$, the number projection operator can be written as an integral in
the complex plane:
\begin{eqnarray}
\hat{P}^{N}=\frac{1}{2\pi i}\oint\frac{z^{\hat{N}}}{z^{N+1}}dz.
\end{eqnarray}
Here we note the property
$\oint\frac{dz}{z^{n}}=2\pi i\delta_{n1}$ with the contour being taken around the origin.
When applied to the BCS wave function of the following form
\begin{eqnarray}
|\Psi\rangle=|BCS\rangle=\prod_{k>0}(\mu_{k}+\nu_{k}\hat{c}_{k}^{\dagger}\hat{c}_{\bar{k}}^{\dagger})|0\rangle,
\end{eqnarray}
one obtains the projected wave function
\begin{eqnarray}
|\Psi_{N}\rangle
=\frac{1}{2\pi i}\oint\frac{d\xi}{\xi^{p+1}}\prod_{k>0}(\mu_{k}+\nu_{k}\xi \hat{c}_{k}^{\dagger}\hat{c}_{\bar{k}}^{\dagger})|0\rangle,
\end{eqnarray}
where we have introduced $\xi=z^{2}$ and used the fact that the pair operator
$\hat{c}_{k}^{\dagger}\hat{c}_{\bar{k}}^{\dagger}$ raises
the particle number by 2, and $p = N/2$ is the number of nucleon pairs. Also we have used the property
$\oint\frac{d\xi}{\xi}=2\pi i$. The integrand in the above equation is a Laurent series in $\xi$. The integration just picks the terms with $\xi^{-1}$, which is the component with $p$ pairs. Using
the fermion anti-commutation relations for the operators $\hat{c}_{k}$ and $\hat{c}_{k}^{\dagger}$, arbitrary matrix elements
can be expressed by the residuals:
\begin{eqnarray}
&&R_{\nu}^{m}(k_{1},\cdots,k_{m})\\\nonumber
&=&\frac{1}{2\pi i}\oint\frac{dz}{z^{(p-\nu)+1}}\prod_{k\neq k_{1},\cdots,k_{m}>0}
(\mu_{k}^{2}+z\nu_{k}^{2}).
\end{eqnarray}
The states listed in the argument of $R(\cdots)$ are to be excluded from the product under the
integral. Suppose that the Hamiltonian of the system has the following form~\cite{P.Ring,A.Bohr} (a single
particle part plus a pure pairing part):
\begin{eqnarray}
\hat{H}&=&\sum_{j>0}\varepsilon_{j}(\hat{c}_{j}^{\dagger}
\hat{c}_{j}+\hat{c}_{\bar{j}}^{\dagger}\hat{c}_{\bar{j}})\\\nonumber
&+&\sum_{j1,j2>0}\bar{v}_{j1,\bar{j1},j2,\bar{j2}}\hat{c}_{j1}^{\dagger}\hat{c}_{\bar{j1}}^{\dagger}\hat{c}_{\bar{j2}}\hat{c}_{j2}.
\end{eqnarray}
The total energy of the system, which is the expectation value of the Hamiltonian, can be
expressed as
\begin{eqnarray}
E^{N}_{proj}
&=& \frac{\langle\Psi_{N}|\hat{H}|\Psi_{N}\rangle}{\langle\Psi_{N}|\Psi_{N}\rangle}\\\nonumber
&=& 2\sum_{j>0}\varepsilon_{j}\nu_{j}^{2}\frac{R_{1}^{1}}{R^{0}_{0}}
+\sum_{j>0}\bar{v}_{j,\bar{j},j,\bar{j}}\nu_{j}^{4}\frac{R_{1}^{1}}{R^{0}_{0}}\\\nonumber
&+&\sum_{j1,j2>0}\bar{v}_{j1,\bar{j1},j2,\bar{j2}}\mu_{j1}\nu_{j1}\mu_{j2}\nu_{j2}\frac{R_{1}^{2}}{R^{0}_{0}}.
\end{eqnarray}
In the second step, we have used the relation $R_{\nu}^{2}(k,k)=R_{1}^{1}(k)$~\cite{P.Ring}. From now on, we adopt a
different notation for the pairing matrix element, i.e.
$G_{j_{1}j_{2}}=-\bar{\nu}_{j_{1}{,}\bar{j_{1}}{,}j_{2}{,}\bar{j_{2}}}$.
 Then the energy of the system can be expressed as
\begin{eqnarray}
E^{N}_{proj}
&=& \sum_{j>0}2[(\varepsilon_{j}-\frac{1}{2}G_{jj}\nu_{j}^{2})\nu_{j}^{2}]\frac{R_{1}^{2}}{R^{0}_{0}}\\\nonumber
&-&\sum_{j1>0}\sum_{j2>0}G_{j1j2}\mu_{j1}\nu_{j1}\mu_{j2}\nu_{j2}\frac{R_{1}^{2}}{R^{0}_{0}}\\\nonumber
&=& \sum_{j>0}2E_{j}\nu_{j}^{2}\frac{R_{1}^{1}}{R^{0}_{0}}-\sum_{j>0}\Delta_{j}\mu_{j}\nu_{j},
\end{eqnarray}
where $\Delta_{j}$ is defined below and we have introduced a new quantity $E_{j}=\varepsilon_{j}-\frac{1}{2}G_{ij}\nu_{j}^{2}$. In the BCS treatment, usually the second term $-\frac{1}{2}G_{ij}\nu_{j}^{2}$ is neglected with the argument that it corresponds only to a renormalization of the single particle energies. In that case $E_{j}$ is simply $\varepsilon_{j}$. This approximation is also adopted in our present work. A variation of the projected energy with respect to $\mu_{j}$ and $\nu_{j}$,
\begin{eqnarray}
(\frac{\partial}{\partial \nu_{j}}-\frac{\nu_{j}}{\mu_{j}}\frac{\partial}{\partial\mu_{j}})E^{N}_{proj}=0,
\end{eqnarray}
leads to the FBCS equation
\begin{eqnarray}
2(\widetilde{\varepsilon_{j}}+\Lambda_{j})\mu_{j}\nu_{j}+\Delta_{j}(\nu_{j}^{2}-\mu_{j}^{2})=0.
\end{eqnarray}
The quantities $\tilde{\varepsilon}_{j}$, $\Lambda_{j}$ and $\Delta_{j}$ are defined as follows:
\begin{eqnarray}
\widetilde{\varepsilon_{j}}&=&(\varepsilon_{j}-G_{jj}\nu_{j}^{2})\frac{R_{1}^{1}}{R^{0}_{0}}\\\nonumber
\Delta_{j}&=&\sum_{k>0}G_{jk}\mu_{k}\nu_{k}\frac{R_{1}^{2}(j,k)}{R^{0}_{0}}(\nu_{j}^{2}-\mu_{j}^{2})\\\nonumber
\Lambda_{j}&=&\sum_{k>0}(\varepsilon_{j}-\frac{1}{2}G_{kk}\nu_{k}^{2})\nu_{k}^{2}\frac{R^{0}_{0}(R_{2}^{2}
-R_{1}^{2})-R_{1}^{1}(R_{1}^{1}-R_{0}^{1})}{(R^{0}_{0})^{2}}\\\nonumber
&-&\frac{1}{2}\sum_{k1,k2>0}G_{k1k2}\mu_{k1}\nu_{k1}\mu_{k2}\nu_{k2}\frac{R^{0}_{0}(R_{2}^{3}-R_{1}^{3})}{(R^{0}_{0})^{2}}\\\nonumber
&+&\frac{1}{2}\sum_{k1,k2>0}G_{k1k2}\mu_{k1}\nu_{k1}\mu_{k2}\nu_{k2}\frac{R_{1}^{2}(R_{1}^{1}-R_{0}^{1})}{(R^{0}_{0})^{2}}.
\end{eqnarray}
The quantity $\Lambda_{j}$ has no counterpart in the conventional BCS equation, where a constant chemical potential
 is chosen to make the expectation value of the number operator equal to the required particle number.
In the derivation of the above equation,
the quantity $\Lambda_{j}$ arises from the differentiation of the residuum integrals with
respect to $\nu_{j}$ and $\mu_{j}$. In the usual BCS theory, $\nu_{j}^{2}$ is the
probability for the pair of states $(j, \bar{j})$ being occupied, and $\mu_{j}^{2}$ is the probability
for this pair of states unoccupied.  In  the FBCS theory,
the corresponding quantities $\epsilon_{j}^{2}$  and $f_j^2$ are:
\begin{eqnarray}
\epsilon_{j}^{2}=\langle\Psi_N|(\hat{c}_{j}\hat{c}_{\bar{j}})^{\dagger}(\hat{c}_{j}\hat{c}_{\bar{j}})|\Psi_{N}\rangle
=\nu_{j}^{2}R_{1}^{1}(j)/R_{0}^{0},
\end{eqnarray}
\begin{eqnarray}
f_{j}^{2}=1-\epsilon_{j}^{2}=\mu_{j}^{2}R_{0}^{1}(j)/R_{0}^{0}.
\end{eqnarray}
To derive the above relations, we have used the recursion relations and derivatives of the
``residuum integrals''~\cite{Dietrich:1964zz}. Of course, the sum of the occupation probabilities is equal to
$N/2$,  i.e., the number of pairs of particles:
\begin{eqnarray}
\sum_{j>0}\epsilon_{j}^{2}=N/2=p.
\end{eqnarray}
 The solutions of the FBCS equation can be formally expressed
as:
\begin{eqnarray}
\mu_{j}^{2}&=&\frac{1}{2}(1-\frac{\widetilde{\varepsilon_{j}}+\Lambda_{j}}{\sqrt{(\widetilde{\varepsilon_{j}}+\Lambda_{j})^{2}+\Delta_{j}^{2}}}),\\\nonumber
\nu_{j}^{2}&=&\frac{1}{2}(1+\frac{\widetilde{\varepsilon_{j}}+\Lambda_{j}}{\sqrt{(\widetilde{\varepsilon_{j}}+\Lambda_{j})^{2}+\Delta_{j}^{2}}}),
\end{eqnarray}
which have the same form as the solutions of the conventional BCS equation, but with $\widetilde{\varepsilon_{j}}$ instead of  $\varepsilon_{j}$.

The total energy in the RMF+BCS model can be simply expressed as
\begin{eqnarray}
E=E_{RMF}+E_{pair}^{p}+E_{pair}^{n},
\end{eqnarray}
with  the pairing energy
\begin{eqnarray}
E_{pair}=-\sum_{k>0}\Delta_{k}u_{k}v_{k}.
\end{eqnarray}
In the usual RMF+BCS model, the densities are determined by the occupation probabilities $v_{i}^{2}$ multiplied with $|\psi_j|^2$, the modulus of
the occupied single particle wave functions.
In the RMF+FBCS model, we merely replace the occupation probabilities $v_{i}^{2}$ by $f_{i}^{2}=v_{i}^{2}R_{1}^{1}/R_{0}^{0}$, i.e.
\begin{eqnarray}
\sum_i v_{i}^{2}\cdots \,\,\,\,\, \Rightarrow \,\,\,\,\,\sum_i f_{i}^{2}\cdots\,\,\,\,\,\,.
\end{eqnarray}

\section{EVALUATION OF THE RESIDUUM INTEGRALS}
To solve the FBCS equation, one needs to  calculate the residuum integrals, i.e.
$R_{0}^{0}$, $R_{0}^{1}$, $R_{1}^{1}$, $R_{1}^{2}$, $R_{2}^{2}$, $R_{1}^{3}$ and $R_{2}^{3}$.
One can simplify the calculations by reducing the number of residuum integrals with several recursion relations.
The first one is
given by Dietrich et al.~\cite{Dietrich:1964zz}, i.e.,
\begin{eqnarray}
R_{\nu}^{m}(k_{1},\cdots,k_{m})&=&R_{\nu+1}^{m+1}(k_{1},\cdots,k_{m},k)v_{k}^{2}\\\nonumber
&+&R_{\nu}^{m+1}(k_{1},\cdots,k_{m},k)u_{k}^{2}.
\end{eqnarray}
With this relation, one third of the total number of independent residuum integrals can be reduced.
Another more powerful relation, firstly deduced by Ma et al.~\cite{PhysRevC.16.1179}, is
\begin{eqnarray}
R_{\nu}^{m}(k_{1},\cdots,k_{m})
&=&\delta_{m\nu}R_{0}^{0}\prod_{i=k_{1},\cdots,k_{m}}\frac{1}{v_{i}^{2}}\\\nonumber
&+&(-1)^{\nu}\sum_{i=k_{1},\cdots,k_{m}}v_{i}^{2(m-\nu-1)}u_{i}^{2\nu}\\\nonumber
&\times&\bigg(\prod_{j=k_{1},\cdots,k_{m}\neq i}\frac{1}{v_{i}^{2}-v_{j}^{2}}\bigg)R_{0}^{1}(i).
\end{eqnarray}

The remaining residuum integrals,
\begin{eqnarray}
R_{\nu}^{m}(k_{1},\cdots,k_{m})&&\\\nonumber
&=&\frac{1}{2\pi i}\oint\frac{dz}{z^{(p-\nu)+1}}\prod_{k\neq k_{1},\cdots,k_{m}>0}(u_{k}^{2}+zv_{k}^{2}),
\end{eqnarray}
can be straightforwardly calculated by replacing $z$ with $r(\cos\theta+i\sin\theta)$, namely,
\begin{eqnarray}
R_{\nu}^{m}(k_{1},\cdots,k_{m})&=&\frac{1}{2\pi i}\oint\frac{r(-\sin\theta+i\cos\theta)d\theta}{[r(\cos\theta+i\sin\theta)]^{(p-\nu)+1}}\\\nonumber
&\times&\prod_{k\neq k_{1},\cdots,k_{m}>0}[u_{k}^{2}+r(\cos\theta+i\sin\theta)v_{k}^{2}].
\end{eqnarray}

\section{GENERAL FEATURES OF THE RMF+FBCS MODEL}
In this section, we  study the general features of the RMF+FBCS model and compare them with those of the conventional RMF+BCS model. For such a purpose, we take the calcium isotopes $^{32-74}$Ca  as examples.
We adopt the commonly used density-independent contact delta interaction $V=-V_{0}\delta(\vec{r}_{1}-\vec{r}_{2})$
for the particle-particle channel in both methods. The only free parameter in the pairing channel
is the pairing strength $V_{0}$, which can be fixed by fitting the pairing gap ($\Delta$) to the experimental odd-even mass difference. The
single particle levels active for the pairing correlation are confined to those within a 10 MeV window around the Fermi surface.

\begin{figure}[tbp]
  \setlength{\abovecaptionskip}{5pt}
  \hspace{-1.1cm}
  \begin{minipage}[c]{0.8\linewidth}
    \centering
    \includegraphics[width=1.2\linewidth]{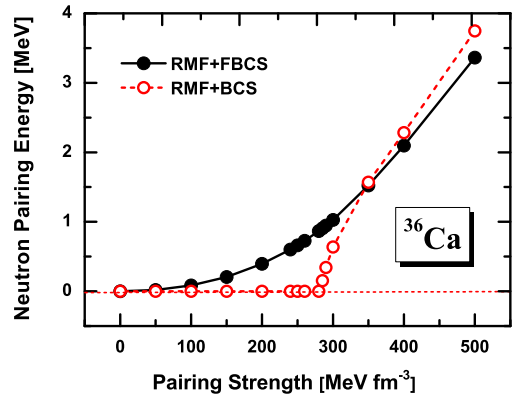}
  \end{minipage}
    \caption{(Color online) Neutron pairing energy of $^{36}$Ca as a function of the pairing strength. \label{fig1}}
\end{figure}

The FBCS method is expected to be able to provide a smooth phase transition
from normal states to superfluid states as a function of the pairing strength.
This is very important because it can show whether the FBCS method can properly describe
the weak pairing limit. In Fig.~\ref{fig1}, the neutron pairing energy of $^{36}$Ca
is plotted as a function of the pairing strength $V_{0}$. Clearly, the FBCS method
does lead to  non-trivial solutions no matter how weak
the pairing strength, while an abrupt transition between superfluid and normal
states arises in the BCS method.
The BCS equation completely fails to give a non-trivial solution
below the critical pairing strength of about 250 MeV fm$^{-3}$.
 Beyond the critical value, the pairing energy in the RMF+BCS model increases rapidly and approaches
that in the RMF+FBCS model in the region of the strong pairing limit($V_{0}=300\sim500$ MeV fm$^{-3}$).
When the pairing strength exceeds 350 MeV fm$^{-3}$, the BCS pairing energy becomes larger than that in the FBCS model,
which can be traced back to the broken gauge symmetry of particle number conservation.

%\begin{figure}[tbp]
 % \setlength{\abovecaptionskip}{5pt}
  %\hspace{-1.1cm}
  %\begin{minipage}[c]{0.9\linewidth}
   % \centering
   % \includegraphics[width=1.2\linewidth]{ca52phasetransition.eps}
  %\end{minipage}
   % \caption{(Color online) Neutron pairing energy of $^{52}Ca$ as a function of the pairing strength. \label{fig2}}
%\end{figure}

\begin{figure}[tbp]
  \setlength{\abovecaptionskip}{5pt}
  \hspace{-1.1cm}
  \begin{minipage}[c]{0.8\linewidth}
    \centering
    \includegraphics[width=1.6\linewidth]{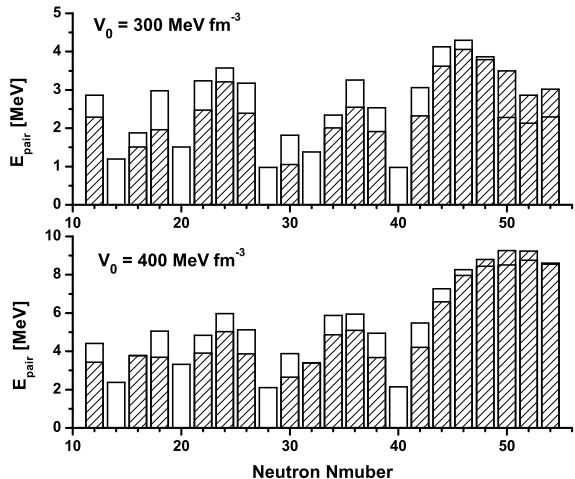}
  \end{minipage}
    \caption{Neutron pairing energies of the calcium isotopes as a function of the neutron number with pairing strengths V$_{0}$=300  MeV fm$^{-3}$ and V$_{0}$=400 MeV fm$^{-3}$, respectively. The results from the RMF+FBCS model (empty columns) are compared with those from the RMF+BCS model (shaded columns). \label{fig2}}
\end{figure}

Now we proceed to study the whole calcium isotopic chain from $^{32}$Ca to $^{74}$Ca.
Two issues of particular interest are the magnitude of the pairing
correlation and how it evolves as a function of the neutron (mass) number.
One can define many different quantities for such a purpose~\cite{P.Ring}.
Here we use the pairing energy defined in Eq.~(17). In Fig.~\ref{fig2}, we compare
the neutron pairing energy of the calcium isotopes $^{32-74}$Ca obtained from
the RMF+FBCS and RMF+BCS calculations with pairing strengths $V_{0}=300$ MeV fm$^{-3}$
and $V_{0}=400$ MeV fm$^{-3}$, respectively. It is clear that the neutron pairing
energies obtained with different pairing strengths show almost the same pattern. Particularly
interesting is that at  $N=14, 20, 28, 32$ and 40, the neutron pairing energy is smaller than  that of their
neighbors in the RMF+FBCS model.
The same scenario occurs in the RMF+BCS model except for $N=32$
with $V_{0}=300$ MeV fm$^{-3}$ where the pairing energy vanishes. This shows that
not only the conventional magic numbers $N=20$, $28$, but also $N=14$, $40$, and to a less extent, $N=32$ shows
some kind of ``magicity'', which seems  to agree with Refs~\cite{PhysRevC.74.021302,nature498,PhysRevC.89.034316,LI201697}.

%\begin{figure}[tbp]
 % \setlength{\abovecaptionskip}{5pt}
 % \hspace{-1.1cm}
 % \begin{minipage}[c]{0.8\linewidth}
 %   \centering
 %   \includegraphics[width=1.2\linewidth]{fig2.eps}
 % \end{minipage}
 %   \caption{(Color online) Neutron pairing energy of isotopic chain of calcium as a function of mass number with pairing strength V$_{0}=300$ and V$_{0}=400$, respectively. \label{fig2}}
%\end{figure}

\begin{figure}[tbp]
  \setlength{\abovecaptionskip}{5pt}
  \hspace{-1.1cm}
  \begin{minipage}[c]{0.8\linewidth}
    \centering
    \includegraphics[width=1.6\linewidth]{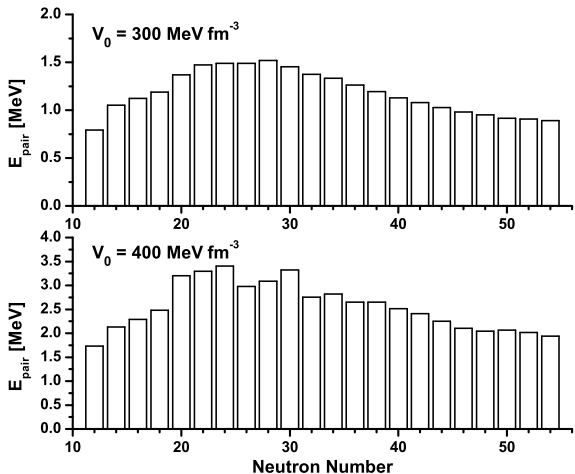}
  \end{minipage}
    \caption{Same as Fig.~\ref{fig2}, but for the proton pairing energies. \label{fig3}}
\end{figure}

In Fig.~\ref{fig3}, we show the proton pairing
energies of the calcium isotopes  as a function of the neutron number.
 It can be seen that the RMF+FBCS pairing energies are still not zero,
even for the proton magic number $Z = 20$, which behave
differently from those in the RMF+BCS calculations.
Furthermore, the proton pairing energies vary lowly as a function of the neutron number,
but the magnitude of this variation is small.

%One should note that the parameters of the
%RMF model are determined with the assumption that there is no pairing correlation in
%magic nuclei. In other words, the magic number effects might be a little bit overestimated
%in the RMF parametrization. Naively speaking, one expects the RMF+FBCS model to
%generate larger (than physical) binding energies for these nuclei.
%Therefore, a proper comparison between the RMF+FBCS calculations and the RMF+BCS calculations
%can only be carried out if the mean-field channel parameters of the RMF+FBCS model
%are readjusted to leave room for further refined treatment of the pairing correlation. Keeping
%this in mind, in the following section, we will study how the present RMF+FBCS model
%can describe nuclear ground-state properties, in comparison with the RMF+BCS
%model.

\section{GROUND-STATE PROPERTIES OF CALCIUM ISOTOPES}
In this section, we study how the bulk ground-state properties of the calcium isotopes can be
described in the RMF+FBCS and RMF+BCS models. The pairing
strength is fixed at $V_{0} = 350$ MeV fm$^{-3}$ in the RMF+BCS model and that
 in the RMF+FBCS model is fixed at $V_{0} = 274$  MeV fm$^{-3}$ by fitting to the odd-even mass differences of the whole
 calcium isotopic chain, defined as the following~\cite{P.Ring,A.Bohr}:
\begin{eqnarray}
\Delta^{(3)}(N,Z)=B(N-1,Z)-2B(N,Z)+B(N+1,Z).
\end{eqnarray}

\begin{figure}[tbp]
  \setlength{\abovecaptionskip}{5pt}
  \hspace{-1.1cm}
  \begin{minipage}[c]{0.8\linewidth}
    \centering
    \includegraphics[width=1.2\linewidth]{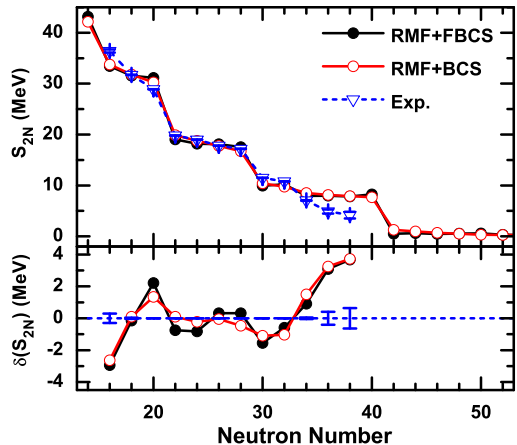}
  \end{minipage}
    \caption{(Color online) Theoretical and experimental~\cite{M.Wang} two-neutron separation energies S$_{2N}$ of the calcium isotopes and
    the difference between them, defined as $\delta(S_{2N})$=$S_{2N}$(th)$-S_{2N}$(exp). \label{fig5}}
\end{figure}

Firstly, we examine the two-neutron separation energy defined as the following:
\begin{eqnarray}
S_{2N}(Z,N)=B(Z,N)-B(Z,N-2),
\end{eqnarray}
where $B(Z,N)$ is the binding energy of a nucleus with proton number $Z$ and neutron
number $N$. In the upper panel of Fig.~\ref{fig5}, the two-neutron separation energies obtained from both models
are compared with their experimental counterparts~\cite{M.Wang}. While in the lower panel, the deviations of the theoretical
two neutron separation energies from their experimental counterparts are shown. One can easily see
that except for $N=36$ and 38, the results of both methods agree with data quite well.
It seems that the $N=40$ magicity effect is overestimated in the RMF model.

\begin{figure}[tbp]
  \setlength{\abovecaptionskip}{5pt}
  \hspace{-1.1cm}
  \begin{minipage}[c]{0.8\linewidth}
    \centering
    \includegraphics[width=1.2\linewidth]{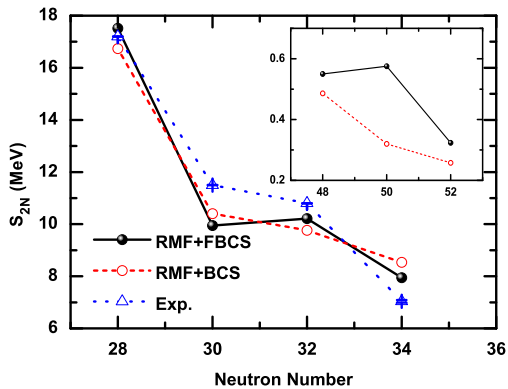}
  \end{minipage}
    \caption{(Color online) Two-neutron separation energies of  $^{48}$Ca, $^{50}$Ca, $^{52}$Ca, and $^{54}$Ca.
    The insert shows those of the   $^{68}$Ca, $^{70}$Ca, and $^{72}$Ca. \label{fig6}}
\end{figure}

A closer look at the two-neutron separation energies of $^{48}$Ca, $^{50}$Ca,
$^{52}$Ca and $^{54}$Ca in Fig.~\ref{fig6} reveals that the experimental sharp drop from $^{52}$Ca to $^{54}$Ca
is better reproduced in the RMF+FBCS model. The same scenario is seen in the inset of Fig.~\ref{fig6} where
 there is a sharp drop from $^{70}$Ca to $^{72}$Ca in the RMF+FBCS model.

\begin{figure}[tbp]
  \setlength{\abovecaptionskip}{5pt}
  \hspace{-1.1cm}
  \begin{minipage}[c]{0.8\linewidth}
    \centering
    \includegraphics[width=1.2\linewidth]{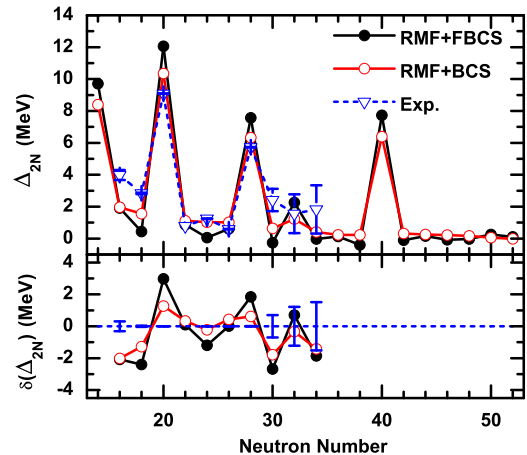}
  \end{minipage}
    \caption{(Color online) Two-neutron gaps of the calcium isotopes  and the differences between
    theoretical and experimental values~\cite{M.Wang} as a function of the neutron number. \label{fig7}}
\end{figure}

 In Ref.~\cite{PhysRevLett.116.152502}, the pairing rotational moment of inertia
 is suggested to be an excellent pairing indicator,
because odd-mass nuclei could contain
the contribution from time-odd fields and better be avoided.
The pairing rotational moment of inertia is proportional to the inverse of
the two-nucleon shell gap indicator $\Delta_{2N}$ \cite{RevModPhys.75.1021}:
\begin{eqnarray}
\Delta_{2N}(Z,N)=2B(Z,N)-B(Z,N+2)-B(Z,N-2)
\end{eqnarray}
In Fig.~\ref{fig7}, the two-neutron shell gaps of the calcium isotopes and the
deviations from their experimental counterparts  are plotted as a function of the
neutron number. It is seen that the RMF+BCS model provides a slightly better description of the experimental
data, especially  for $^{40}$Ca and $^{48}$Ca. This can be easily understood from
the definition of $\Delta_{2N}$.  In the BCS  method the pairing correlation is
only effective on open-shell nuclei and reduces the two-neutron shell gaps of magic nuclei (compared with
pure mean field models or the FBCS method).

From the studies of the two-neutron separation energies and two-neutron gaps of the calcium
isotopes, it seems that the RMF+BCS calculations are of similar
quality or even slightly better than the RMF+FBCS calculations. This finding is not
surprising. It is closely related to how we obtained the RMF parameters.
The NL3 RMF parameterization is fitted to the ground-state properties of 10 magic or even-even nuclei~\cite{PhysRevC.55.540}.
That is to say, from the very beginning, we only expect the residual pairing
correlation to make open-shell nuclei more bound but leave closed-shell nuclei
unchanged. The BCS and Bogoliubov methods are perfect candidates to achieve
this as we can easily see in Fig.~\ref{fig3}, though they break the gauge symmetry of
particle number. In contrary, the FBCS method makes closed-shell nuclei more
bound than what the BCS or Bogoliubov method does and leaves open-shell nuclei
more or less unchanged. Therefore, it is quite natural that no significant
improvement has been observed. To really appreciate the FBCS method, in particular to
improve the agreement with the experimental data, the mean-field effective force has to be
readjusted to leave room for incorporating these higher-order correlations \cite{Heenen:2001bt}.
Due to the present strategy used to fit the RMF parameters, at least part of the
pairing effect for magic nuclei has been compensated by artificially large magic
number effects at the order of several MeV

%\begin{figure}[tbp]
%  \setlength{\abovecaptionskip}{5pt}
%  \hspace{-1.1cm}
%  \begin{minipage}[c]{0.8\linewidth}
%    \centering
%    \includegraphics[width=1.2\linewidth]{fig6.eps}
%  \end{minipage}
%    \caption{(Color online) The potential energy surface of N=32 and 34 ($^{52}$Ca, $^{54}$Ca) as a function of the quadrupole deformation parameter. The pairing strength,V$_{0}=$350, is identical for FBCS and BCS. \label{fig6}}
%\end{figure}

In addition to the binding energies and related quantities, one can study the root mean square (r.m.s.) radii as well as the deformations of the calcium isotopes.
We found that they turn out to be similar in both the RMF+BCS and RMF+FBCS models and therefore refrain from showing them explicitly.
On the other hand, we notice that close to the neutron drip line $N\ge 50$, the r.m.s. radii in the RMF+BCS model are
slightly larger than those in the RMF+FBCS model, at the order of 0.05 fm. However, because of the harmonic oscillator basis adopted, we do not expect that either of our methods can properly describe the r.m.s. radii or the density distributions close to the neutron drip line.  Nevertheless, we notice that the RMF+BCS and RMF+FBCS models can sometime change the occupation probability of certain single particle levels close to the Fermi surface, and thus modify the density distributions. When the continuum states are more properly treated, this may have some impact on the spatial distributions of drip line nuclei. To illustrate this point, we investigate $^{54}$Ca in detail below.

\begin{figure}[tbp]
  \setlength{\abovecaptionskip}{5pt}
  \hspace{-1.1cm}
  \begin{minipage}[c]{0.8\linewidth}
    \centering
    \includegraphics[width=1.2\linewidth]{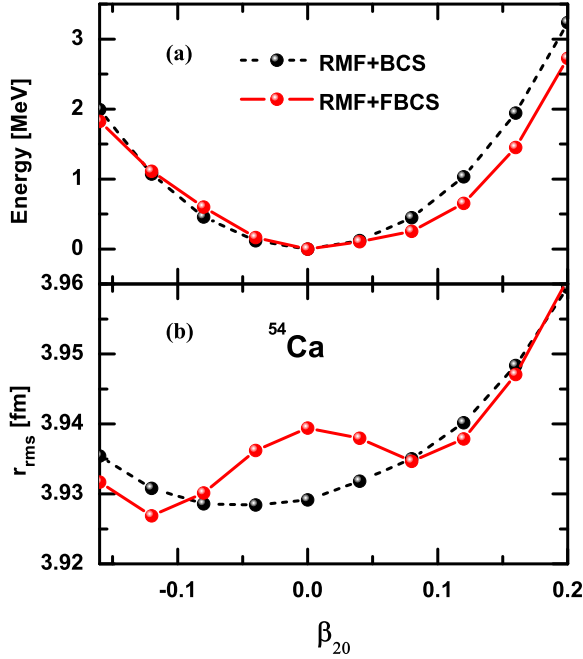}
  \end{minipage}
    \caption{(Color online) Potential energy surface and root mean square radius of $^{54}$Ca as a function of the deformation parameter $\beta_{20}$ obtained in the RMF+BCS (dashed line)  model and the  RMF+FBCS (solid line) model.  \label{fig8}}
\end{figure}

In the upper panel of Fig.~\ref{fig8}, we plot the potential energy surface of $^{54}$Ca as
a function of the quadrupole deformation parameter $\beta_{20}$.
The curves obtained in the two models look quite similar, both yielding a minimum
at $\beta_{20}=0$, but the RMF+FBCS energy at large deformations becomes larger.
In the lower panel of Fig.~\ref{fig8}, the neutron r.m.s. radius of $^{54}$Ca is also shown as
a function of $\beta_{20}$. Surprisingly, we see a bump developed in the center of the RMF+FBCS curve, different from  the RMF+BCS case.~\footnote{We notice that increasing the pairing strength in the RMF+FBCS model will
reduce the bump a little bit but the structure remains even for a pairing strength of 400 MeV fm$^{-3}$.  In addition, the appearance of
such a phenomenon also depends on the adopted mean-field parameters.}

\begin{figure}[tbp]
    \centering
    \includegraphics[width=0.5\textwidth]{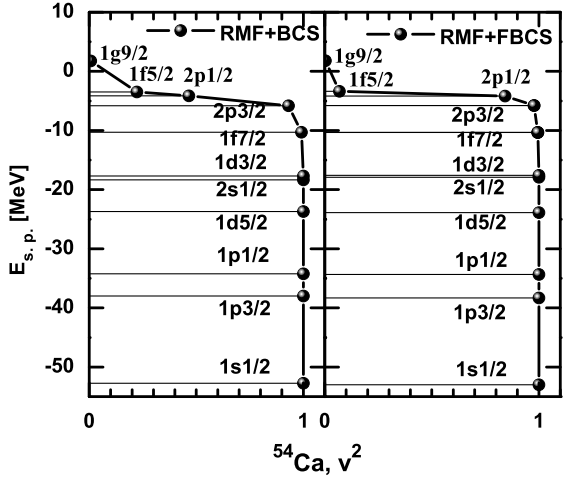}
    \caption{(Color online) Occupation probabilities of the neutron single particle levels of $^{54}$Ca obtained in
    the RMF+BCS model (left) and the RMF+FBCS model (right). \label{fig10}}
\end{figure}

 Since the binding energy  at $\beta_{20}=0$ is similar to each other, such a difference can only originate from the different occupation probabilities of the single particle states
close to the Fermi surface. This is indeed the case as shown in Fig.~\ref{fig10}. We see that the occupation
probability of the neutron $2p_{1/2}$ state in the RMF+ FBCS is much larger than that of
the RMF+BCS model. In the latter, more particles are scattered to the neutron $1f_{5/2}$ orbit.
This explains why at $\beta_{20}=0$, the RMF+BCS and RMF+FBCS models predict a similar
binding energy, but a different neutron r.m.s. radius.

\begin{figure}[htpb]
 \centering
    \includegraphics[width=0.5\textwidth]{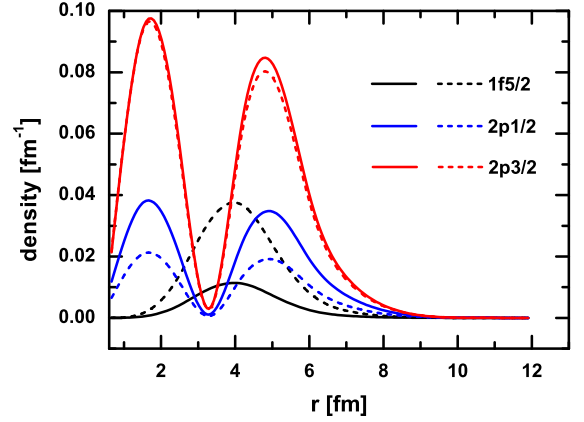}
    \caption{(Color online) Neutron density of the  $1f_{5/2}$,  $2p_{1/2}$, and $2p_{3/2}$ orbits of $^{54}$Ca
     calculated in the RMF+FBCS (solid lines) and RMF+BCS (dashed lines) models.  \label{fig11}}
\end{figure}

In Fig.~\ref{fig11}, we plot the
density distributions of the  neutron1f$_{5/2}$, 2p$_{1/2}$ and
2p$_{3/2}$ orbits. The Nilsson  quantum numbers
are those of the dominant component in the expansion of the wave
function in terms of the axial harmonic oscillator basis. Clearly, in the two methods,
the relative contributions from the $2p_{1/2}$ and $1f_{5/2}$ orbits are quite different.
In the RMF+FBCS model, the contribution from the $2p_{1/2}$ orbit, which extends farther away from the center,
is larger than that from the $1f_{5/2}$ orbit.
While in the RMF+BCS model, the opposite is true.
These are the reasons behind the seemingly unusual behavior observed in Fig.~\ref{fig10}.

\section{SUMMARY}
We have formulated a particle number conserving BCS method, the so-called
FBCS method, in the relativistic mean field model. It
is shown the RMF+FBCS model can properly describe the weak pairing limit.
A detailed study of the calcium isotopes reveals that the RMF+FBCS results for
the two-neutron separation energies and two-neutron gaps are similar to those of the RMF+BCS calculations;
and also the density distributions are roughly the
same in both calculations (therefore not shown). Overall we do not find essential improvement in
the description of the ground state properties of the calcium isotopes.

On the other hand, we notice that the neutron r.m.s radii at the neutron drip lines can
be somewhat larger in the RMF+BCS model than in the RMF+FBCS model. In addition, our study showed that the FBCS method can change
the occupation probability of certain single particle orbitals around the Fermi surface and
therefore affect the neutron r.m.s radius. For the case of $^{54}$Ca, the increase
of the radius is only about 0.02 fm, but this can be larger for more neutron rich nuclei
with similar configurations.
However, due to the incorrect asymptotic behavior of the harmonic oscillator wave
functions, the expansion in a localized HO basis is not appropriate for the description
of drip line nuclei~\cite{PhysRevC.68.034323}, particular for their density distributions. To treat the continuum more properly, one may solve the RMF model in
coordinate space~\cite{PhysRevC.68.054323,Meng19983} or adopt the
Woods-Saxon basis ~\cite{PhysRevC.68.034323,PhysRevC.82.011301}. Implementing a
particle number conserving BCS approach or Bogoliubov approach in such models and study its impact on drip line nuclei is
of great interest both experimentally and theoretically. Such works are in progress.

%\begin{figure}[tbp]
%  \setlength{\abovecaptionskip}{5pt}
%  \hspace{-1.1cm}
%  \begin{minipage}[c]{0.8\linewidth}
%    \centering
%    \includegraphics[width=1.7\linewidth]{ca54radius.eps}
%  \end{minipage}
%    \caption{The neutron mean square radius $R_{n}$ as a function of quadrupole deformation obtained by the RMF+FBCS (filled circle) and the RMF+BCS (open circle) calculations for $^{54}$Ca, respectively. The delta pairing force is employed, where V$_{0}$=274.5 Mev fm$^{-3}$ for FBCS and V$_{0}$=350.0 Mev fm$^{-3}$ for BCS. \label{fig1}}
%\end{figure}

\begin{acknowledgments}
An Rong and Shi-Sheng Zhang acknowledge valuable discussions with Prof. Shan-Gui Zhou of Institute of Theoretical Physics, Chinese Academy of Sciences.
This work is supported in part by the National Natural Science Foundation of China under Grants No. 11522539, No. 11735003, No. 11775014 and No.11375022.

\end{acknowledgments}

\bibliography{refs}

\end{document}